\documentstyle[11pt,newpasp,twoside,epsf]{article}
\def\edcomment#1{\iffalse\marginpar{\raggedright\sl#1\/}\else\relax\fi}
\reversemarginpar

\begin{document}
\title{Modelling the non--thermal emission from galaxy clusters}
%
 \author{Gianfranco Brunetti}
\affil{Istituto di Radioastronomia del CNR\\ 
via P. Gobetti, 101, I-40129, Bologna (ITALY)}

\begin{abstract}
We discuss the relevant processes for the relativistic
electrons in the ICM and
the possible mechanisms responsible for the production of
these electrons.
We focus on the origin of the radio halos giving some of the
observational diagnostics which may help in discriminating among
the different models proposed so far.
Finally, we briefly discuss the discrepancy between the
value of the magnetic field assuming an inverse Compton (IC)
origin of the hard X--ray emission (HXR) and that obtained
from Faraday Rotation Measurements (RM).
\end{abstract}

\section{Introduction}

The most important evidence for relativistic electrons in clusters 
of galaxies comes from the diffuse synchrotron radio emission
observed in about 35 \% of the clusters selected with X--ray
luminosity $> 10^{45}$erg s$^{-1}$ (e.g., Giovannini, 
Tordi, Feretti, 1999; Giovannini \& Feretti 2001).
The diffuse emissions are referred to as radio halos and/or radio
mini--halos when they appear projected on the center of the
cluster, while they are called relics when they are found in the
cluster periphery.

The difficulty in explaining the extended radio halos arises
from the combination of their $\sim$Mpc size, and the short
radiative lifetime of the radio emitting electrons.
Indeed, the diffusion time necessary to the radio 
electrons to cover such distances is orders of magnitude
greater than their radiative lifetime.
To solve this problem Jaffe (1977) proposed continuous 
in situ reacceleration of the relativistic electrons.
The in situ reacceleration scenario 
was quantitatively reconsidered by Schlickeiser
et al.(1987) who successfully reproduce the 
integrated radio spectrum of the Coma halo.
In the framework of the in situ reacceleration model,
Harris et al.(1980) first suggested that 
cluster mergers may provide the energetics
necessary to reaccelerate the relativistic particles.
The role of mergers in particle acceleration and 
in the amplification of the magnetic fields was than 
investigated in more detail by De Young (1992) and
Tribble (1993).
Alternatively, to avoid the energy loss problem 
of the radio electrons, Dennison (1980) first 
suggested that the radio emission in radio halos 
may be emitted by a population of secondary electrons 
continuously injected by hadronic interactions.
{\it Primary} and the {\it secondary}  
electron models still constitute the basis
of the more recent theoretical works 
developed on the argument (see Sect.3).
Additional evidence for the presence of non--thermal phenomena
in clusters of galaxies comes from the detection in a number of
cases of EUV excess emission (e.g., Bowyer et al. 1996;
Lieu et al., 1996; Bergh\"ofer et al., 2000; Bonamente et al., 2001)
and of HXR excess emission in the case of the Coma cluster
and A2256 (Fusco--Femiano et al., 1999, 2000;
Rephaeli et al., 1999; Rephaeli \& Gruber, 2002;
Fusco--Femiano, this meeting).
While, with the exception of the Coma and of the Virgo
clusters, the EUV detections are still
controversial (e.g., Bergh\"ofer et al., 2000; 
Bergh\"ofer, Bowyer, Nevalainen, this meeting),
the HXR detections are quite robust as they are
independently obtained
by different groups and with different X--ray observatories
(BeppoSAX, RXTE).
On the other hand, while there is agreement on the 
IC origin at least of the most significant cases of
EUV excess emission
(e.g., Hwang 1997; Bowyer \&
Berg\"ofer 1998; Ensslin \& Bierman 1998;
Sarazin \& Lieu 1998;
Atoyan \& V\"olk 2000; 
Brunetti et al. 2001b, Petrosian 2001, Tsay et al., 2002), 
the origin of the XHR is still debated.
The XHR excess may be generated by IC scattering of relativistic
electrons off the CMB photons
(Fusco--Femiano et al., 1999,2000;
Rephaeli et al., 1999;
V\"olk \& Atoyan 1999; Brunetti et al.2001a;
Petrosian 2001; Fujita \& Sarazin 2001).
Alternatively XHR might also result from bremsstrahlung
emission from a population of supra--thermal
electrons (e.g., Ensslin, Lieu, Biermann 1999;
Blasi 2000; Dogiel 2000; Sarazin \& Kempner 2000).
Both the IC model and the bremsstrahlung
interpretation have problems :
the first one would require cluster magnetic field strengths
smaller than that inferred from RM observations
(e.g., Clarke, Kronberg, B\"oringer 2001),
the second one would require a too large amount of
energy to mantain a substantial fraction of the
thermal electrons
far from the thermal equilibrium for more
than $10^8$yrs (e.g., Petrosian, 2001).
In this contribution we will describe the 
populations of relativistic electrons expected in clusters
of galaxies and we will focus
on the origin of radio halos and HXR emission
from galaxy clusters.
We refer to the contribution by Ensslin for the origin
of radio relics and to the contributions by Bowyer for
a review on the EUV excesses.

\section{Competing mechanisms at work}

Before discussing the origin of the emitting electrons
in galaxy clusters, it is convenient to briefly review
the basic processes which modify
the energy and spectrum of the relativistic
electrons.
It is important to underline that the efficiency of
these processes is related to the energy of the electrons
in a way that depends on the particular process so
that the time evolution
of electrons with different energies
is dominated by different processes.
More specifically relativistic 
electrons with energy $mc^2 \gamma$
in the intracluster medium (ICM) lose energy via two main
processes:

a) ionization losses and Coulomb collisions :

\begin{equation}
\left( {{ d \gamma }\over{d t}}\right)_{\rm ion}^-=-
1.2 \times 10^{-12} n_{\rm th}
\left[1+ {{ {\rm ln}(\gamma/{n_{\rm th}} ) }\over{
75 }} \right]
\label{ion}
\end{equation}

\noindent
where $n_{\rm th}$ 
is the number density of the thermal plasma.

b) synchrotron and IC radiation :

\begin{equation}
\left( {{ d \gamma }\over{d t}}\right)_{\rm syn+ic}^-=-
1.3 \times 10^{-20} \gamma^2 
\left[ \left( {{ B_{\mu G} }\over{
3.2}} \right)^2 {{ \sin^2\theta}\over{2/3}}
+ (1+z)^4 \right]
\label{syn+ic}
\end{equation}

\noindent
where $B_{\mu G}$ is the magnetic field strength in
$\mu G$, and
$\theta$ is the pitch angle of the emitting electrons.

On the other hand, the electrons in the ICM
can be re--accelerated by several mechanisms. 
Two relevant cases in clusters of galaxies
are shock acceleration and acceleration via
wave -- particle interaction 
(e.g., MHD or HD turbulence).

\noindent
{\it a) shock acceleration}
(e.g., Blandford \& Eichler, 1987) yields an energy gain :
%

\begin{equation}
\left( {{ d \gamma }\over{d t}}\right)_{\rm sh}^+
\simeq \gamma {{ {U_-}^2 }\over{f}} \left( {{ f-1}\over{ f+1}}
\right) {1 \over {3 {\cal K}(\gamma) }}
\label{shok}
\end{equation}

\noindent
${{\cal K}}(\gamma)$ is the spatial diffusion coefficient, 
$U_-$ is the velocity of the plasma in the
region before the shock discontinuity (measured in the
shock frame and in unit of $c$), and
$f$ is the
shock compression ratio.

\noindent
{\it b) Acceleration via turbulence}:
if the resonance scattering condition 
is satisfied (e.g. Hamilton \& Petrosian 1992),
turbulent Alfven waves can accelerate electrons via resonant
pitch angle scattering.
A power law energy spectrum of the Alfven waves:

\begin{equation}
P(k)=
b {{ B^2}\over{8 \pi}}
{{ s-1}\over{k_o}}
\left(
{{ k \over {k_o} }}
\right)^{-s}
\label{aspectrum}
\end{equation}

\noindent
in the range $k_o < k < k_{\rm max}$ is assumed, where
$k$ is the wave number
($k_o << k_{\rm max}$) and $b$ is
a normalization factor indicating the
fraction of the energy density of the
magnetic field $B$ in energy of waves.
Under this assumption
it can be shown that the systematic energy gain is
(e.g., Blasi 2000; Ohno et al. 2002):

\begin{equation}
\left( {{ d \gamma }\over{d t}}\right)_{\rm A-tur}^+
\sim  b { {\pi} \over c}
\left(
1 - {1\over s}
\right)
v_{\rm A}^2
\left(
{{ e B }\over{m_{\rm e} c^2 }}
\right)^{2-s}
k_o^{s-1}
\times
\gamma^{s-1}
\label{ono1}
\end{equation}

\noindent
where $v_{\rm A}$ is the Alfven velocity.

\noindent
MHD turbulence can also accelerate relativistic particles
in radio sources
via Fermi--like processes (e.g., Lacombe 1977; Ferrari et al. 1979).
Under the simple assumption of a quasi--monochromatic
turbulent scale responsible for particle acceleration
(e.g., Gitti, Brunetti, Setti 2002)
the systematic energy gain is :

\begin{equation}
\left( {{ d \gamma }\over{d t}}\right)_{\rm F-tur}^+
\simeq 4 \times 10^{-11} \gamma
{{ {v_{\rm A}}^2 }\over{ l }}
\left( {{\delta B}\over{B}} \right)^2
\label{tur}
\end{equation}

\noindent
where $l$ is the
distance between two peaks of turbulence and
$\delta B/B$ is the fluctuation in a peak
of the field intensity with
respect to the average field strength.
Note that Eq.(6) and Eq.(5) have the same dependence on
the relevant parameters for $P(k) \propto k^{-2}$.

A comparison between losses and gain (shocks, or turbulence with
$s=2$) processes is given in Fig.1a,b:
radio emitting electrons ($\gamma \sim 10^4$)
have radiative lifetimes of $\sim 10^8$yrs, in addition 
the acceleration of electrons with $\gamma < 10$ and 
$\gamma > 10^5$ appears 
extremely difficult (see also Petrosian, this
meeting).

\begin{figure}
\plottwo{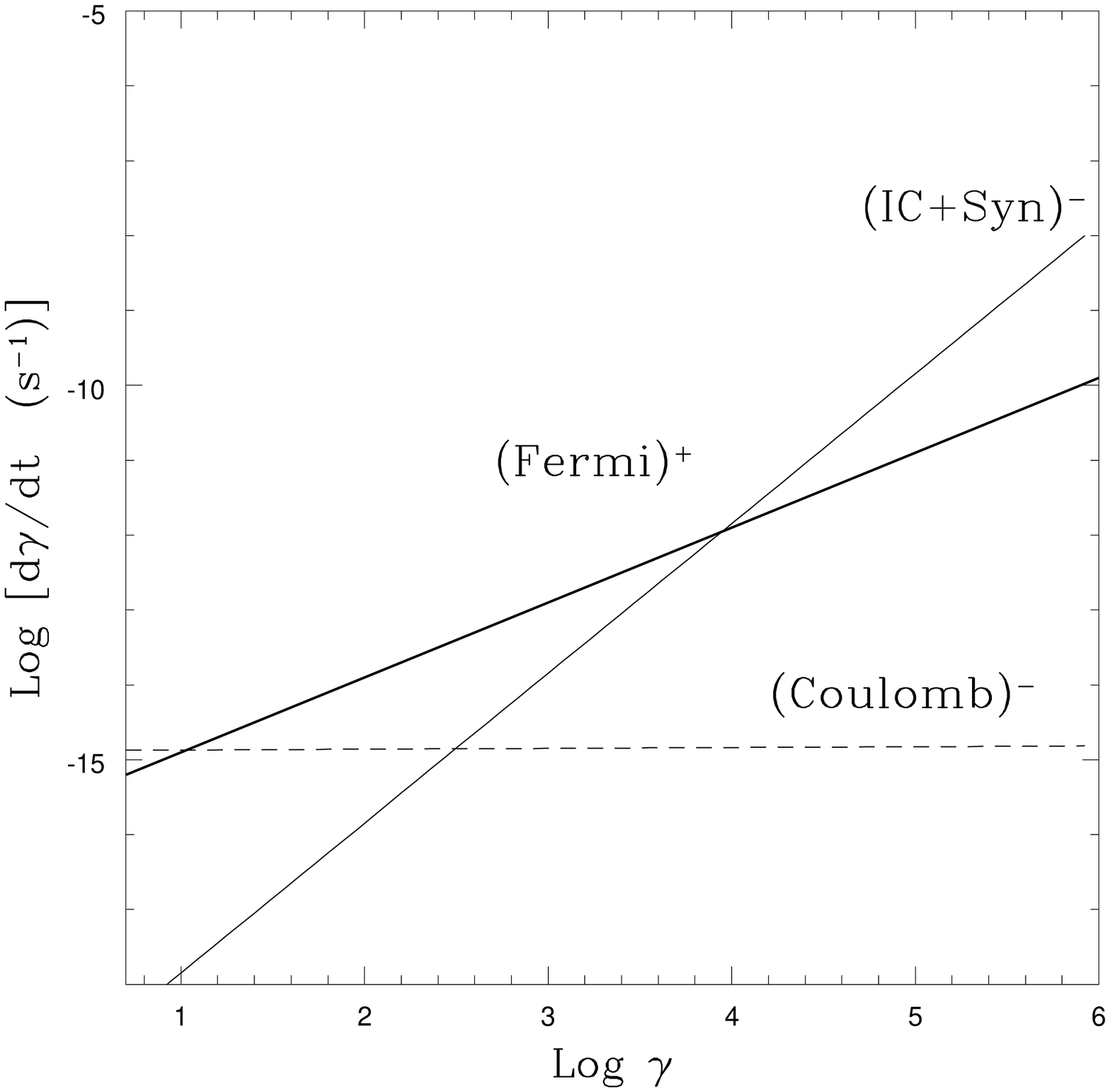}{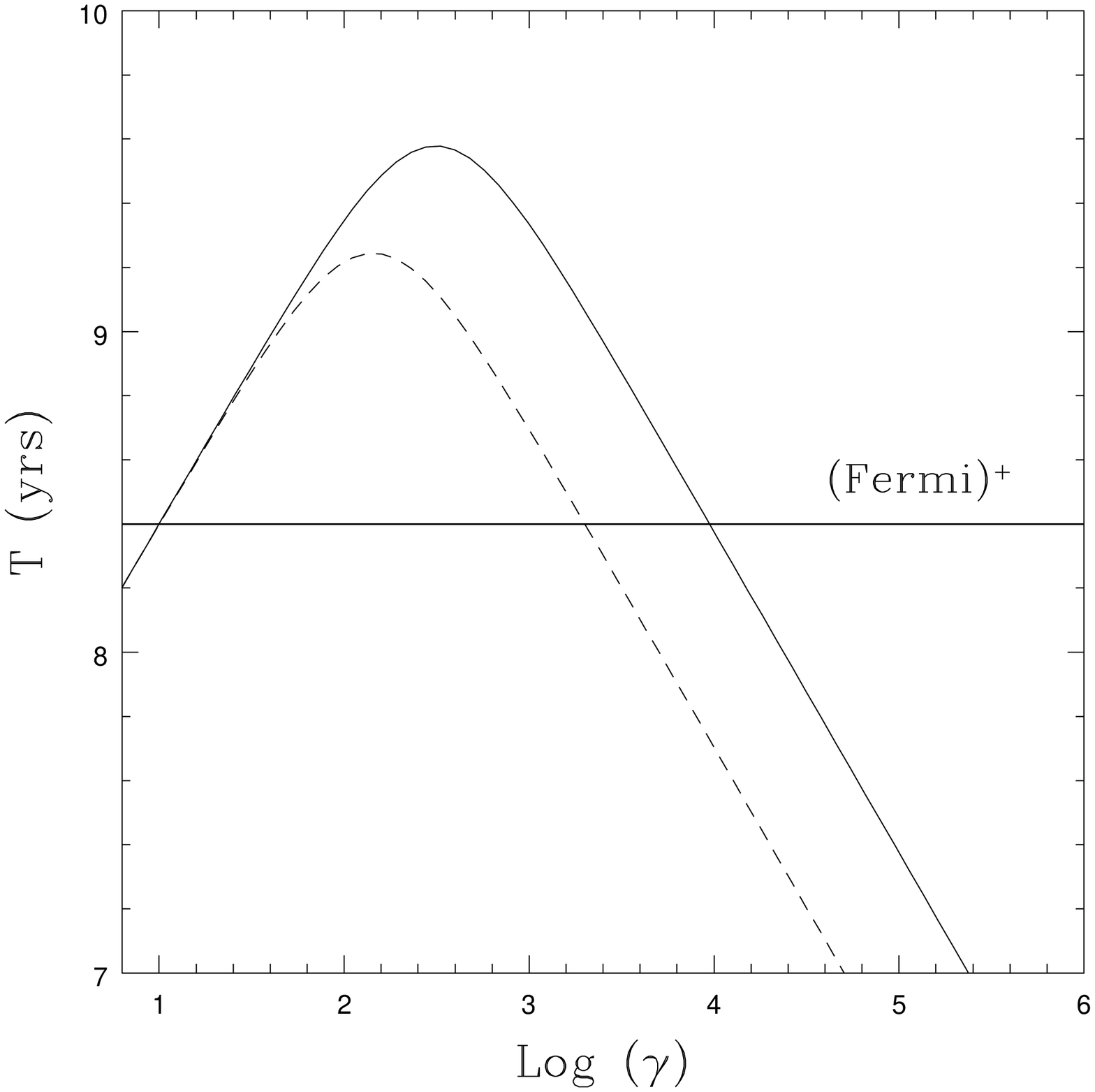}
\caption{
{\bf Panel a) }: The efficiency of Fermi
acceleration
(an acceleration time $\sim 2.5 \times 10^8$ yrs is assumed,
solid thick line), IC+syn losses (solid line),
and Coulomb losses (dashed line) are reported for typical
conditions in the ICM as a
function of $\gamma$ of the 
electrons. 
{\bf Panel b) }: The radiative lifetime of 
electrons, $\tau_{\rm loss} = \gamma / (d\gamma/dt)$, 
is reported
as a function of $\gamma$ of the electrons.
The calculations are performed for $n_{\rm th}=10^{-3}$
cm$^{-3}$, $B=1 \mu G$, assuming $z=0$ (solid line) and
$z=0.5$ (dashed line).
Electrons with a radiative lifetime larger than the
acceleration time 
($2.5 \times 10^8$ yrs) can be accelerated.}
\end{figure}

\section{Electron populations in galaxy clusters}

It has been shown that a magnetic field with strength 
$\geq 0.1 \mu$G can easily store the bulk of cosmic rays
in the cluster volume for a time greater 
than the Hubble time (e.g., Berezinski, Blasi and Ptuskin 1997).
If this holds in the case of protons, the confinment is 
much easily obtained for electrons.
Indeed the diffusion length of the particles decreases with
the energy, and thus it is much shorter in
the case of relativistic electrons than in that of 
the relativistic protons.

Here we focus our attention on the case of 
the populations of relativistic electrons. 
Electrons can be injected in the ICM by
different processes :

\begin{itemize}
\item[{\it i)}] {\bf Acceleration by shocks (Pop.A)}:
Radio observations of supernovae indicate
that strong shocks convert at least a few percent of their 
energy into the acceleration of relativistic particles.
Thus, merger shocks may represent 
a natural acceleration mechanism for the relativistic 
electrons in galaxy clusters (e.g., Sarazin, 1999).
Particle acceleration by merger shocks
has been studied in detail (e.g., Takizawa \& Naito, 2000;
Miniati et al., 2001; Fujita \& Sarazin, 2001).
and it might explain the
apparent correlation between the non--thermal 
phenomena and the presence of merger activity in
clusters of galaxies (e.g., Buote, 2001 and ref. therein).

\item[{\it ii)}] {\bf Reaccelerated electrons (Pop.B)}:
There is a number of sources of relativistic electrons in
galaxy clusters.
In particular active galaxies (e.g., radio galaxies), 
merger shocks, supernova and galactic winds
can efficiently inject relativistic protons and
electrons in the cluster volume over cosmological time 
(e.g., Sarazin, 2002 and ref. therein).
High relativistic electrons have very short radiative lifetimes,
however, when they reach energies of $\gamma \sim 100-300$ 
they survive for some billion years (Fig.1b)
and thus
can be accumulated in the cluster volume.
Cluster mergers may produce a significant level 
of turbulence in the ICM, in this case 
Alfven waves and/or some other Fermi--like processes
diffuse in the cluster volume 
could reaccelerate $\gamma \sim 100-300$ relativistic
electrons to the higher energies required to explain
radio halos (e.g., Brunetti et al., 2001a).
Electron reacceleration has been also invoked for the origin
of radio mini--halos in cooling flow clusters.
In this case the energy for the reacceleration may be 
provided by the cooling flow itself (Gitti et al.2002).

\item[{\it iii)}] {\bf Secondary electrons (Pop.C)}:
Dennison (1980) first pointed out that a possible source
of the relativistic electrons in radio halos is the
continuous injection due to the decay of charged mesons
generated in cosmic ray ion collisions.
This idea has been reconsidered in detail
in the model by Blasi \& Colafrancesco (1999), and  
then by Dolag \& Ensslin (2000) assuming a 
radial profile of the cluster magnetic field
taken from numerical simulations. 
In order to explain the connection between cluster
mergers and radio halos, Ensslin (1999) proposed that 
relativistic protons are released from radio ghosts into 
the ICM during a cluster merger event.
More recently, Miniati et al. (2001)
have developed numerical simulations
of cluster formation and 
calculated the injection of primary relativistic
protons by strong shocks.
These authors find that, under some assumptions,
the resulting secondary electrons might produce 
diffuse synchrotron emission in agreement with
some of the observed properties of radio halos.

\end{itemize}

\noindent
From a theoretical point of view 
the above electron populations are very reasonable, 
thus it may very well be that all of them
contribute to the injection of the relativistic
electrons in the ICM.
In addition it should be noticed that the final 
electron population may be due to 
a complicated mix of all the
above reported processes. For example, shocks may accelerate
relic electrons contributing to the Pop.B, but they can also
(re)accelerate relativistic protons increasing the
rate of injection of secondary electrons (Pop.C).
In addition to the acceleration of relic electrons
(Pop.B), cluster turbulence would (re)accelerate protons 
increasing the rate of injection of secondary electrons (Pop.C).
Finally, secondary electrons can be (re)accelerated by 
shocks and/or cluster turbulence powered by a merger
event (Pop.A,B).

\section{Possible diagnostics}

A possibility to break the degeneracy on the origin
of the emitting electrons might be provided 
by some observational
diagnostics.

\subsection{Diffusion lengths}

The $\sim$ Mpc size of most radio halos suggests that
a population of electrons accelerated by shocks in the
clusters (Pop. A) cannot significantly contribute to the 
observed diffuse radio emission.
This is because, after being accelerated by a shock, 
the synchrotron emitting electrons have a short
($\sim 10^8$yrs) radiative lifetime and cannot
diffuse over the cluster volume.

\begin{figure}
\plottwo{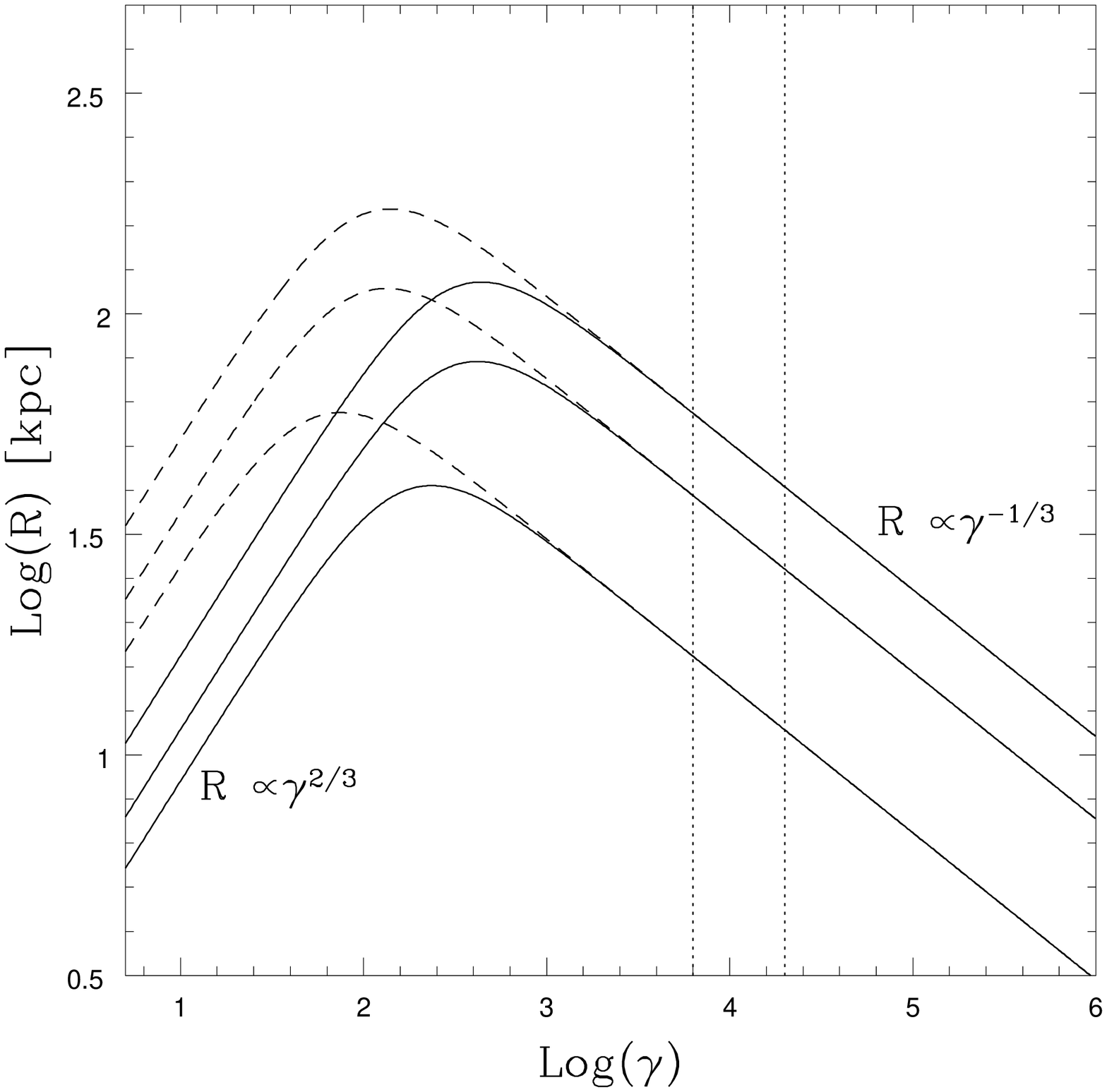}{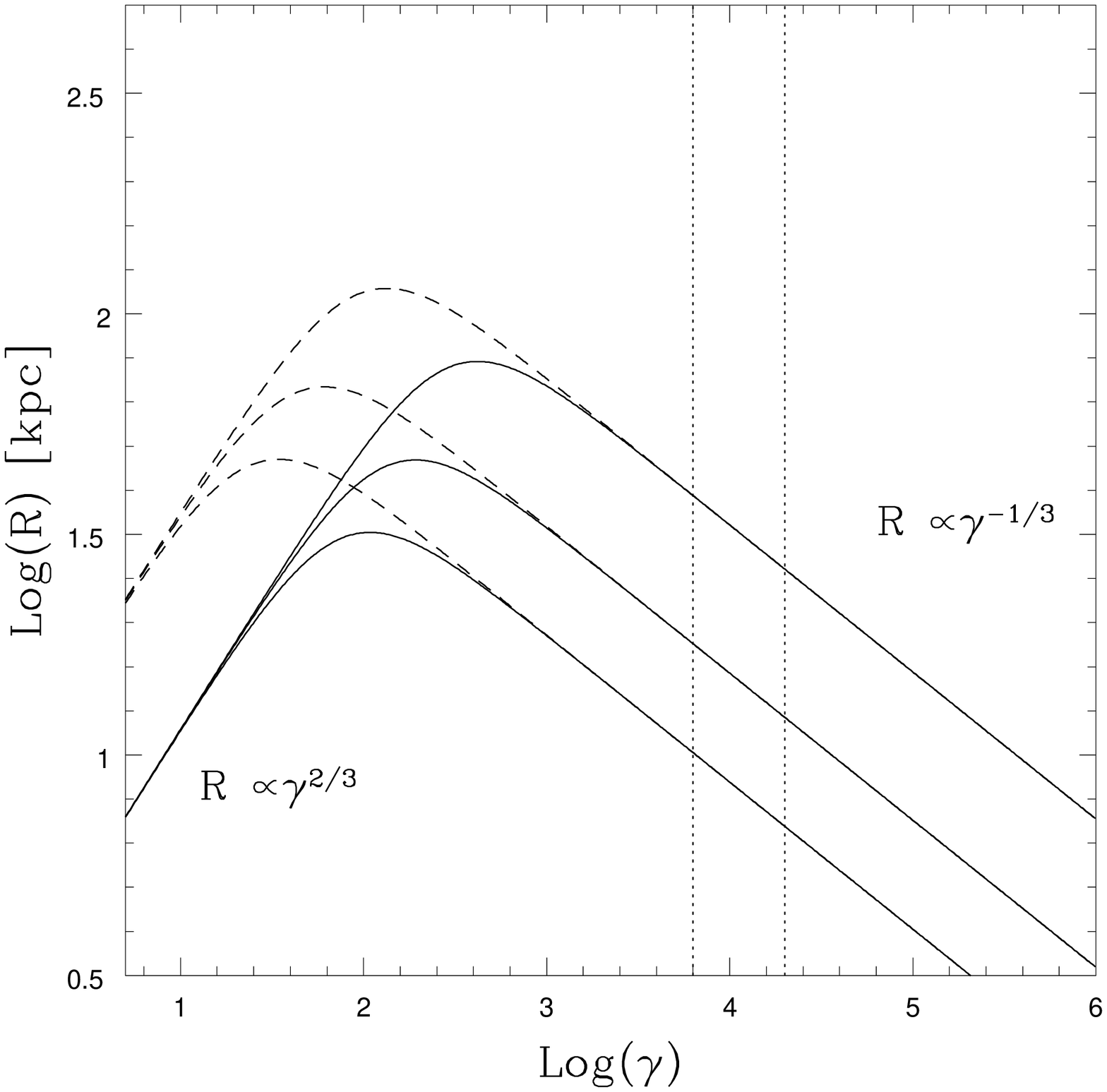}
\caption{
{\bf Panel a) } : Diffusion lengths are reported
as a function of $\gamma$ of the electrons.
Calculations are performed at $z=0$,
for $n_{\rm th} = 10^{-3}$
cm$^{-3}$ (solid lines) and $10^{-4}$
cm$^{-3}$ (dashed lines) assuming (from the bottom) 
$B$=5, 1, and 0.1 $\mu G$.
{\bf Panel b) } : Diffusion lengths as in panel a), 
but assuming $B=1 \mu G$, and
$z=0$, 0.5, and 1.0 (from the top).
The energy range of the radio emitting electrons
is reported in both panels (vertical dotted lines).
}
\end{figure}

In order to better quantify the typical diffusion length 
$R_{\rm d}$ of
the radio electrons, we should derive the diffusion time 
$\tau_{\rm diff} \sim R_{\rm d}^2/(4 {{\cal K(\gamma)}})$, 
%
where ${\cal K}(\gamma)$ is the spatial diffusion coefficient.
We assume a Kolmogorov spectrum of the
magnetic field (e.g., Blasi \& Colafrancesco, 1999) and obtain :

\begin{equation}
{\cal K(\gamma)} \simeq
1.8 \times 10^{28} L_{20 \rm kpc}^{2/3}
\left( {{ \gamma }\over{ B_{\mu G}}} \right)^{1/3}
\end{equation}

\noindent
where $L_{20 \rm kpc}$ is the largest coherence scale 
of the field and $B_{\mu G}$ is the magnetic field
strength in $\mu G$.
The diffusion length $R_{\rm d}$ is obtained when the diffusion time 
equals the radiative lifetime, $\tau_{\rm loss}=\gamma (d\gamma/dt)^{-1}$
(Sect.2). 
%
In the most interesting regime, when IC losses dominate
(i.e., for $\gamma >> 10^3$ and $B_{\mu G} < 3$), we
obtain:

\begin{equation}
R_{\rm d} ({\rm kpc}) \sim
36 (1+z)^{-2}
\left(
{{ \gamma }\over
{10^4}} \right)^{-1/3}
\left( 
{{
L_{\rm kpc} }\over{
B_{\mu G}^{1/2} }
}
\right)^{1/3}
\end{equation}

\noindent
A more general result is reported in Fig.2 : the maximum diffusion
length is obtained for electrons with $\gamma \sim 100-1000$, while
radio emitting electrons cannot diffuse for more than about 50 kpc.

Another possibility is given by a scenario in which a fast
shock crosses the cluster center and accelerates the
relativistic electrons across the cluster volume.
In this case the diffusion time of the electrons is replaced
by the crossing time of the shock that should be shorther than
the radiative lifetime of the radio emitting electrons, i.e. :

\begin{equation}
\tau_{\rm cross}({\rm yrs})
\sim
{{ 10^9 D({\rm Mpc}) }\over{ U_{\rm sh}/10^3 }}
< \tau_{\rm loss} (\gamma\sim 10^4)
\end{equation}
\begin{figure}
\plottwo{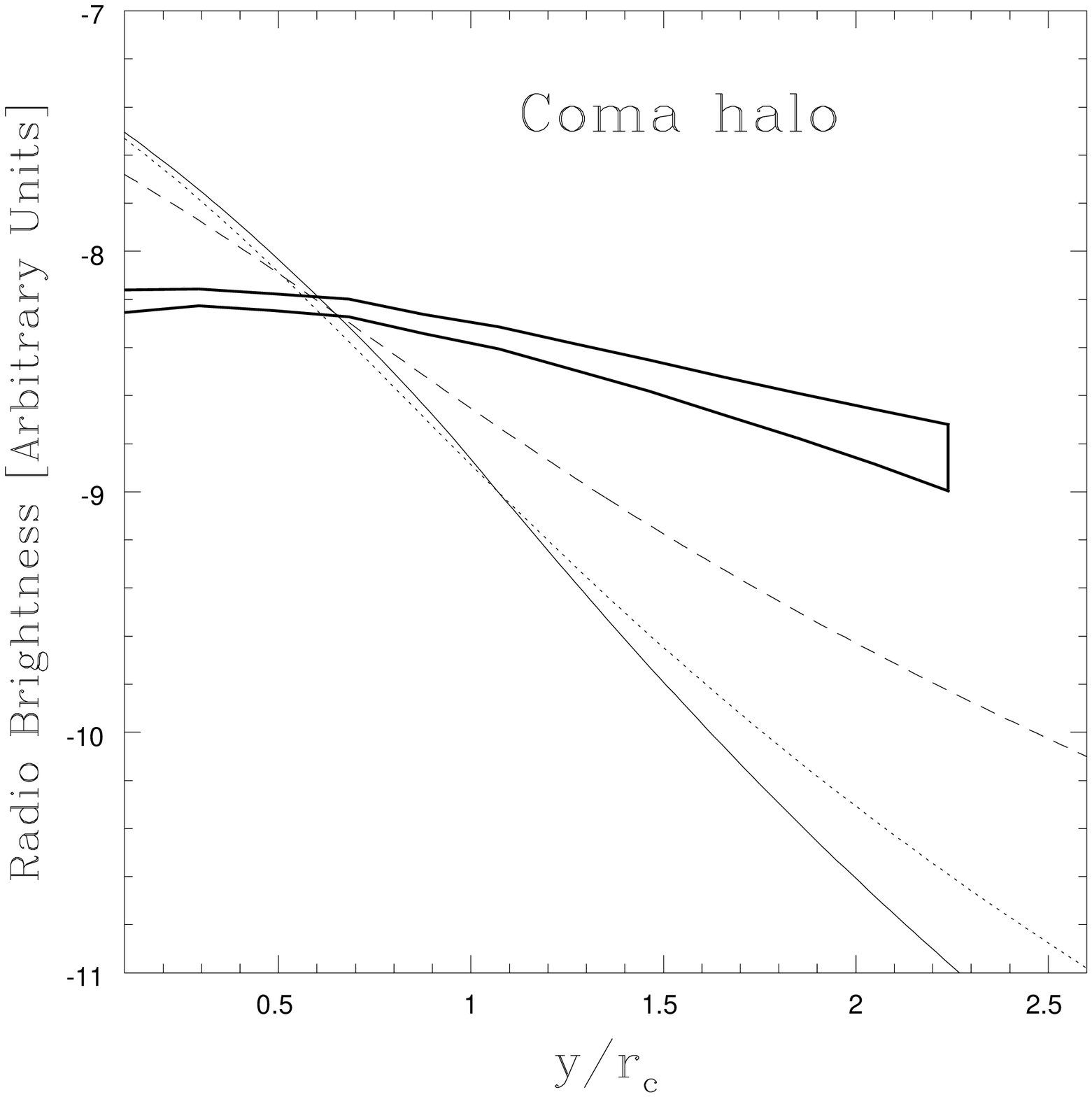}{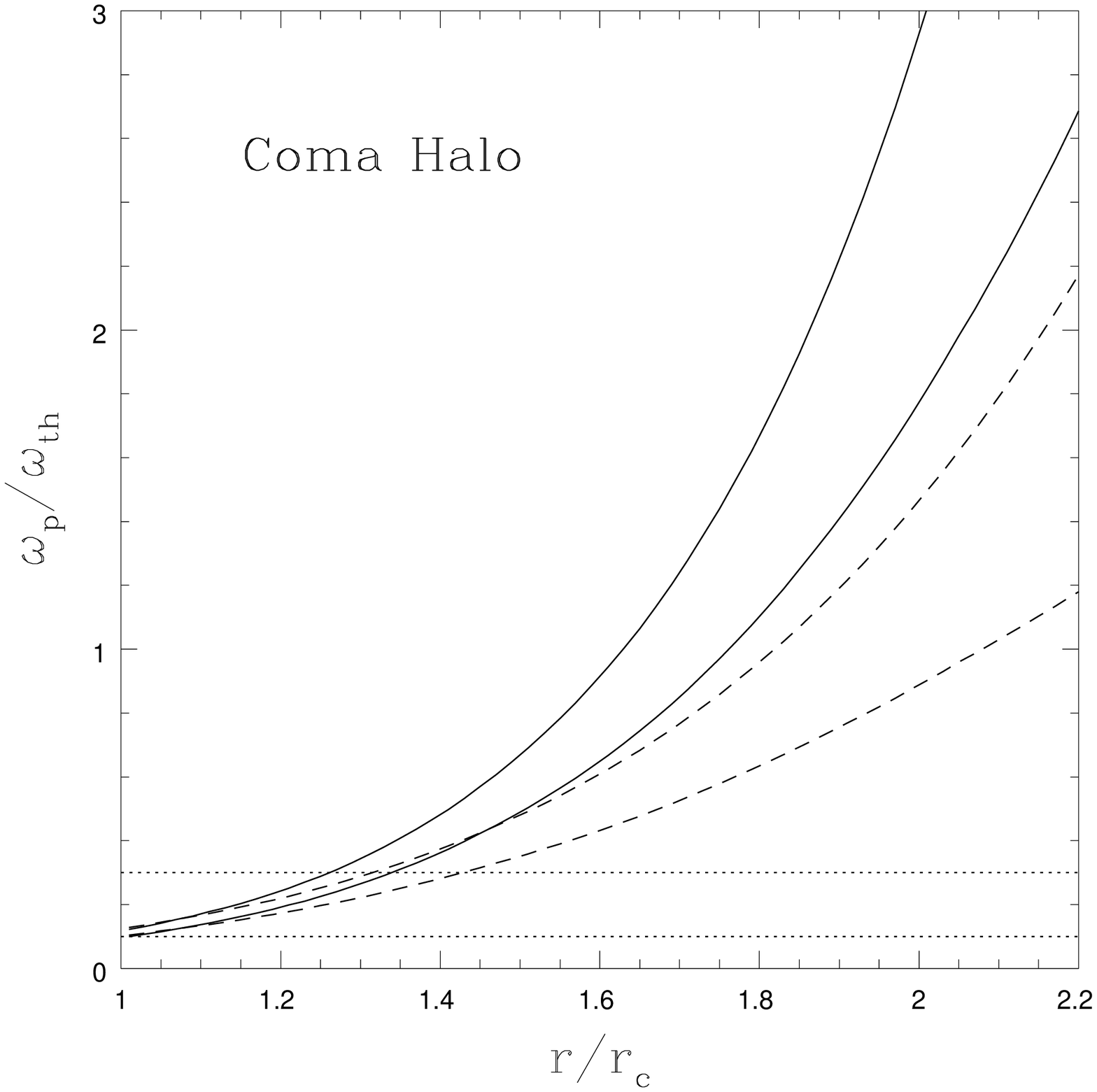}
\caption{
{\bf Panel a) }: Synchrotron brightness profiles calculated
in the case of the Coma cluster due to secondary models
are reported as a function of the projected distance from
the center (in units of the core radius).
A central magnetic field $B=2.0 \mu G$
is assumed.
The calculations assume: ${\cal F}=$const. with
a profile of $B$ from freezing approximation (dotted line)
and from numerical simulations (solid line), and
a constant energy density of the relativistic protons
with $B$ from freezing approximation (dashed line).
The observed brightness profile
(thick lines) is taken from Govoni et al. 2001.
{\bf Panel b) }: The ratio between relativistic and
thermal proton energy density required to match the data
is reported as a function of distance from the center.
Calculations are done for the dotted and solid line models
in Panel a).}
\end{figure}

\noindent
which for a typical total extension of $\sim 2$ Mpc would
require an unreasonaby high 
Mack number of the shock ${\cal M}_{\rm sh} > 5$. 
A detailed calculation of the electron population accelerated
by merger shocks has been recently
developed by Miniati et al.(2001).
These authors calculate the resulting 
synchrotron radiation and they find indeed  
that the morphology of the resulting 
emission is similar to that of radio relics
rather than to radio halos.
Furthermore, due to the field compression in the shock, 
the emitted synchrotron radiation is highly
polarized (as in the case of radio relic)
in contrast with the very low polarization found in
the radio halos.
Additional evidence against a direct
link between merger shock acceleration and non--thermal
emission has been obtained
by Gabici \& Blasi (2002) who have shown that the 
low Mach number expected in the merger shocks would accelerate
a very steep spectrum of protons and electrons which cannot
account for the observed synchrotron spectrum of radio halos
(also in the case of secondary models).

\subsection{Radio brightness profiles}

The timescale of the
p--p collision, which is the process responsible
for the injection of secondary electrons (Pop C),  
is $\propto n_{\rm th}^{-1}$.
Consequently, for a given number density of the relativistic
protons, secondary electrons are expected to be injected 
in the denser regions and 
the radio emission would be stronger 
in the cluster core.
So far, a quantitative comparison of the observed radio and 
thermal bremsstrahlung
X--ray profiles ($b_{\rm r}$ and $b_{\rm x}$,
respectively) was obtained for 5 extended radio halos
(Govoni et al. 2001; Feretti et al. 2001).
In two cases a linear correlation exists between
the radio and X--ray brightness (i.e.,
$b_{\rm r} \propto b_{\rm x}^b$, with
$b\sim 1$) whereas in the remaining three cases 
a sub--linear trend is found (Coma: $b$=0.64, A2319: $b$=0.82,
A2163: $b$=0.64).
On the other hand, simple - but viable - secondary models
would predict $b >1$ trend (Dolag \& Ensslin 2000).
In this Section we explore if this discrepancy 
can be accomodated. 
We assume a power low energy distribution of
the injected relativistic protons:

\begin{equation}
N_{\rm p}(\epsilon,R) =
N_{\rm p}(R) \epsilon^{-s}
\end{equation}

\noindent
where $N_{\rm p}(R)$ gives the spatial distribution of the protons.
In this case, following standard recipes for the calculation
of proton proton decay (e.g., Mannheim \&
Schlickeiser, 1994) and assuming 
time independent conditions (e.g., Dolag \& Enslin 2000),
it can be shown that the energy distribution of the 
relativistic electrons is also a power law :

\begin{equation}
N_{\rm e}(\epsilon,R) =
C_{\rm e}
{{
n_{\rm th}(R) N_{\rm p}(R) }\over{
B^2(R) + B_{\rm cmb}^2}}
\epsilon^{-\delta}
\end{equation}

\noindent
where $C_{\rm e}$ is a constant, and
the slope of the electron spectrum is
$\delta={4\over 3} s + {1\over 3}$.
In order to calculate the number density of the secondary 
electrons we parameterize the number density of the 
relativistic protons with that of the thermal plasma:

\begin{equation}
N_{\rm p}(R) = C_{\rm p} n_{\rm th}(R) {\cal F}(R) kT
\end{equation}

\noindent
where $C_{\rm p}$ is a constant, and 
${\cal F}(R)$ gives the ratio between relativistic and thermal protons
energy densities.
The synchrotron brightness profile is thus
given by:

\begin{equation}
b_{\rm syn}(y) = C_{\rm e} C_{\rm p}
C_{\rm syn} \nu^{-\alpha} kT \times
\int_{y} {{ dR \, R }\over{ \sqrt{ R^2 -y^2}}}
n_{\rm th}^2(R) {\cal F}(R)
{{
{B(R)}^{1+\alpha} }\over{
B^2(R) + B_{\rm cmb}^2}}
\end{equation}

\noindent
We remind that the thermal brightness emission 
from a cluster is given by :

\begin{equation}
b_{\rm th}(y) \propto
\int_{y} {{ dR \, R }\over{ \sqrt{ R^2 -y^2}}}
n_{\rm th}^2(R) \Lambda(T)
\end{equation}

\noindent
and thus the ratio between thermal and synchrotron 
brightness depends on the
quantity $\Phi =
{\cal F}(R) B(R)^{1+\alpha}/( B^2(R) + B_{\rm cmb}^2 )$
in Eq.(14), which is $\propto {\cal F}(R) B(R)^{1+\alpha}$
for a magnetic field strength $< 3 \mu$G.
More specifically, if $\Phi(R)$ is a decreasing function
of $R$ the synchrotron profile will be narrower than the X--ray
thermal profile.
In Fig.3a we report the comparison between the synchrotron
brightness
profile from secondary models and the observed one.
The theoretical profiles are considerably steeper than
observed. This happens 
in the case of both flux freezing approximation
($B \propto n_{\rm th}^{2/3}$) and of
a radial dependence of the field as that from 
numerical MHD simulations (Dolag, Bartelmann and Lesch 2002). 
These simulations suggest a rapid decrease of the field in 
the regions out of the cluster core ($B \propto n_{\rm th}$).
In order to reproduce the observed brightness profile
with secondary models, ${\cal F}(R)$ is then forced to rapidly 
increase with $R$ (Fig.3b).
As a net result, we find that secondary models can reproduce 
the profile of the Coma halo 
only by forcing the energy density of the
relativistic protons to be considerably larger than that of
the thermal ICM out of the cluster radius; this is quite
unreasonable.
On the other hand, Miniati et al.(2001) 
have shown that secondary models can account for the
radio extension of a Coma like halo.
We stress that our calculations are not in contradiction with
Miniati et al.(2001) as we have assumed a physical model for
the magnetic field strength profile, whereas those authors assumed
a value of the magnetic field ($\sim 3 \mu$G) 
roughly constant up to several core radii from the cluster center.
In this case the resulting synchrotron emissivity
at $> r_{\rm c}$ would obviously be increased considerably
with respect to our calculations.

\subsection{Integrated and radial spectral steepenings}

Radio observations of the best studied radio
halo, Coma C, have discovered a cut-off around 1 GHz 
in the integrated synchrotron spectrum 
(e.g., Deiss et al. 1997) and 
a strong steepening of the spectrum 
with increasing
the distance from the center
(Giovannini et al. 1993; Fig.4a).
The presence of a similar radial spectral
steepening has been also found in the well studied 
radio mini--halo of the Perseus cluster (Sijbring, 1993; 
Fig.4b).
So far, the lack of similar multifrequency radio data 
for other radio halos (or mini--halos)
does not allow to understand if the 
synchrotron spectral properties of Coma and Perseus 
are common among radio halos and mini--halos.

\noindent
Brunetti et al.(1999, 2001a) pointed out that 
synchrotron radial spectral steepenings are
expected in the case of extended radio halos
if the synchrotron radiation is emitted by 
reaccelerated electrons (Pop B).
In these models, the maximum energy of the reaccelerated 
electrons is expected at
$\gamma_{\rm c} \sim 2.5 \times 10^4/
(\tau_{\rm acc}/10^8{\rm yrs})$, 
where $\tau_{\rm acc}$ is
the reacceleration time.
Since the
Fermi II--like processes in the ICM
give typical acceleration 
time scales $\tau_{\rm acc} > 10^7$yrs
(e.g., Eilek \& Weatherall 1999),  
it is $\gamma_{\rm c} < 10^5$.
Consequently, a cut--off in the synchroyron spectrum
might be present in the radio band.
Another consequence of these models is that, if
the field strength in clusters
decreases with distance from the center 
(e.g. Dolag et al. 2002), 
the corresponding frequency of the cut--off in the
synchrotron spectrum should decrease with the distance 
from the center yielding a possible radial steepening of
the spectrum between two fixed frequencies
(see also Kuo et al., this proceedings).
\begin{figure}
\plottwo{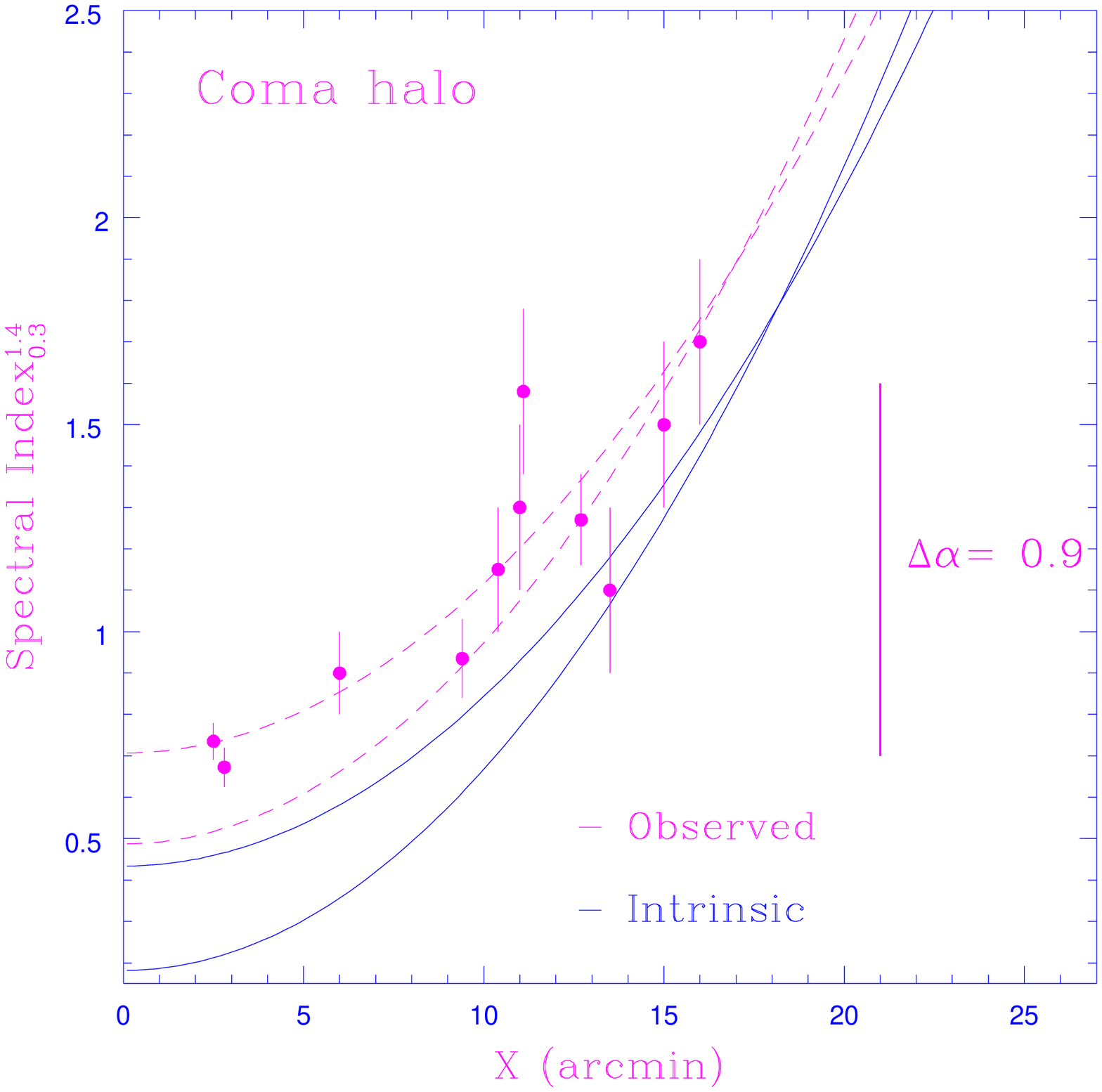}{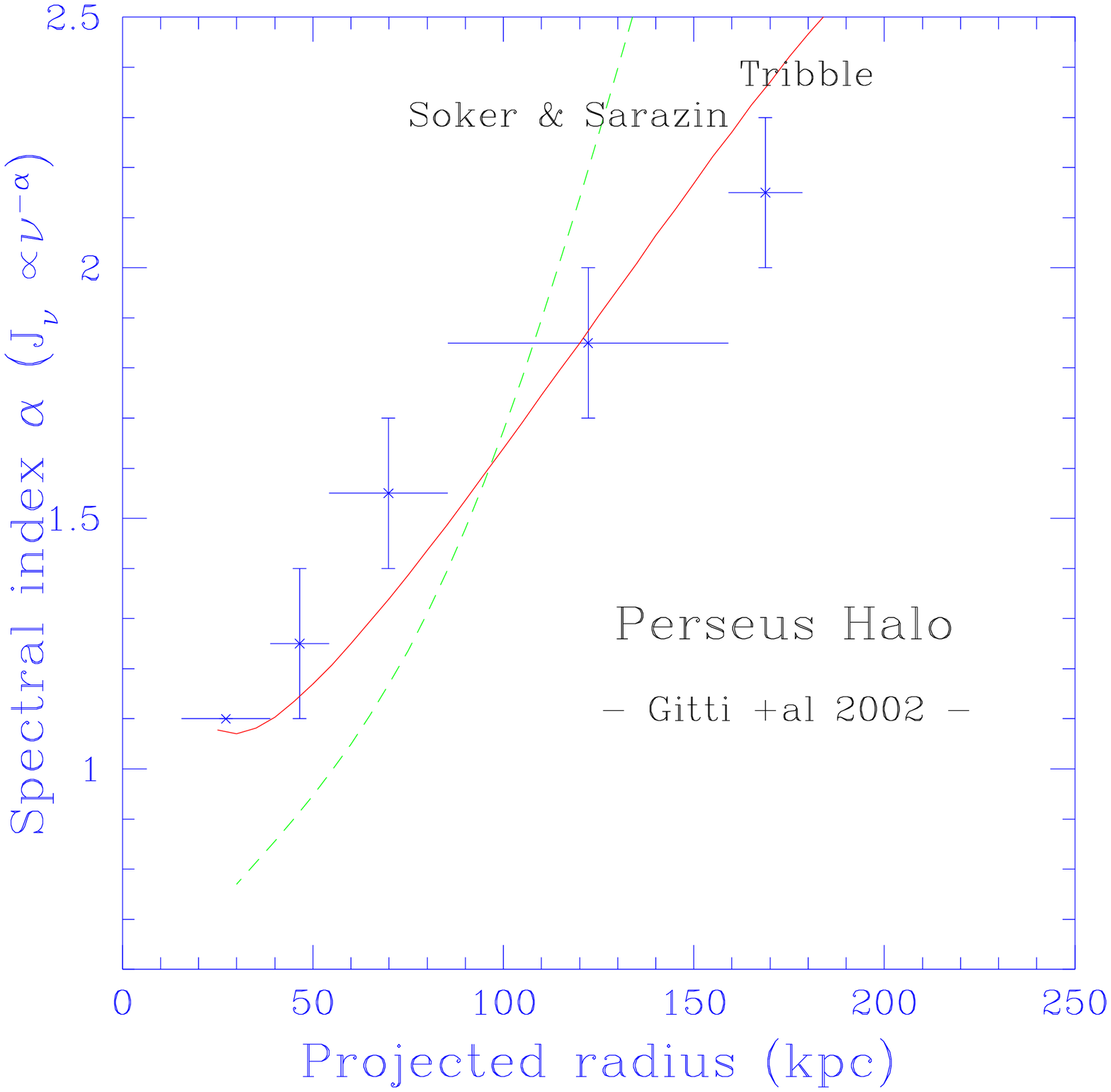}
\caption{
Synchrotron spectral index of the Coma radio halo
({\bf Panel a)} and of the mini--halo in the Perseus
cluster ({\bf Panel b)} are reported as a function
of the distance from the center.
{\bf Panel a)}: data taken from Giovannini et al. 1993,
models from Brunetti et al. 2001a.
{\bf Panel b)}: data taken from Sijbring 1993, models
from Gitti et al. 2002.
}
\end{figure}
On the other hand, as already discussed, if the spectrum of the
primary protons is a power law, the secondary
electron-positron pairs (Pop C) from $\pi^{\pm}$
decay have also a power low spectrum.
This is still true ($\gamma > 1000$), if the complete 
proton--proton cross section is taken into 
account in the calculations
(e.g., Blasi 2001) and 
the spectrum ($\nu > 10$ MHz) of the synchrotron emission 
predicted by secondary models should be a straight power law 
without a cut--off.
One possibility to obtain a cut--off in the spectrum of the
secondary electrons, and thus in the emitted synchrotron
spectrum, is to assume a cut--off in the energy distribution 
of the primary relativistic protons. This should be
at $E_{\rm p} \leq 50$ GeV to have the cut--off of the
spectrum from the secondary electrons at the $\sim$GHz.
Although, at present, such a cut--off cannot be 
ruled out by direct observations, 
we stress that it is very unlikely as it is in contrast with
the observations of the spectrum of (Galactic) cosmic rays 
and with the theoretical expectations from the most 
accepted acceleration mechanisms.
Cosmic ray protons in the Galaxy are detected up to 
$E_{\rm p} \geq 10^{20}$ GeV and independently
on their origin, no cut--off 
is observed at least up to 
$E_{\rm p}\sim 4 \times 10^6$ GeV.
So far, all the mechanisms
invoked to inject the primary population of
relativistic protons in galaxy clusters do not
predict a cut--off in the spectrum of the protons at
low energies.
In particular, merger shocks are expected to 
accelerate protons up to $E_{\rm p}\sim 10^7-10^9$ GeV
(e.g., Blasi, 2001), AGNs might accelerate relativistic
protons in jets and hot spots at $E_{\rm p} \geq 10^8$ GeV
(e.g., Biermann, 1995) and
SNRs, which - indeed -
are likely to produce at least the Galactic cosmic rays
up to the observed
{\it knee} at $E_{\rm p} \sim 4\times 10^6$ GeV, 
can accelerate protons at 
$E_{\rm p} \sim 10^3-10^8$ GeV
(e.g., Bhattacharjee \&
Sigl, 2000).
Studies aimed at constraining the spectrum of
secondary electrons in clusters and their contribution
to the radio halos are important 
(Blasi, Brunetti, Gabici, 2002, in prep.).

\section{Hard X--ray emission and the $B$--field discrepancy}

In this Section we focus on the IC model,
in particular the aim of this Section is to critically
review the discrepancy
between the magnetic field strength as requested
by the IC interpretation and that
from RM observations (e.g., Carilli \& Taylor, 2001 and ref.
therein).
The measure of the magnetic field strengths with the
IC method and via RM observations
clearly involves different spatial averages of the magnetic
field itself : the first measure provides a volume average
of the field on scales $\geq$Mpc, while the second one
provides a weighted average of the field vector and thermal
gas density along the line of sight.
In addition, the IC method is very sensitive to 
the spectrum of the relativistic electrons especially
in the case of $B < 1 \mu G$.
Taking into account the radial dependence of the 
thermal gas and magnetic field strength, 
Goldshmidt \& Rephaeli (1993) first showed that the 
field strength estimated with the IC method is expected 
to be smaller than that `measured' with the RM observations.
In addition, as shown in Fig.5a,
the presence of a high energy cut--off in
the spectrum of the emitting electrons (e.g., Brunetti et al.
2001a; Fujita \& Sarazin 2001) might further increase the
discrepancy of the field strengths obtained making use
of the two methods.
Indeed, the ratio between the typical Lorentz factor of the
radio emitting electrons and of those emitting IC radiation
in the HXR band is
$\gamma_{\rm syn}/\gamma_{\rm HXR} 
\sim 3.5 \times B^{-1/2}_{\mu G}$, thus, in the case $B < 1 \mu G$, 
a cut--off close to $\gamma_{\rm syn}$ would reduce the 
synchrotron emission without affecting the IC in the HXR band.
As a net result, 
the ratio between radio synchrotron and IC HXR flux is reduced.
Since the IC magnetic field is derived by such a ratio, 
the argument can be reversed so that given an observed
ratio between radio and HXR flux, the assumption of 
a cut--off in the electron spectrum (close to the energy of the radio 
electrons) allows us to obtain a
value of the magnetic field strength higher than that calculated
with the standard power law assumption.
This effect, combined with possible anisotropies in the 
pitch angle distribution of the emitting electrons and with observative
biases (Petrosian, 2001), may alleviate 
the discrepancy between the magnetic field values 
as obtained assuming
an IC origin of the HXR and those as estimated by RM observations.
\begin{figure}
\plottwo{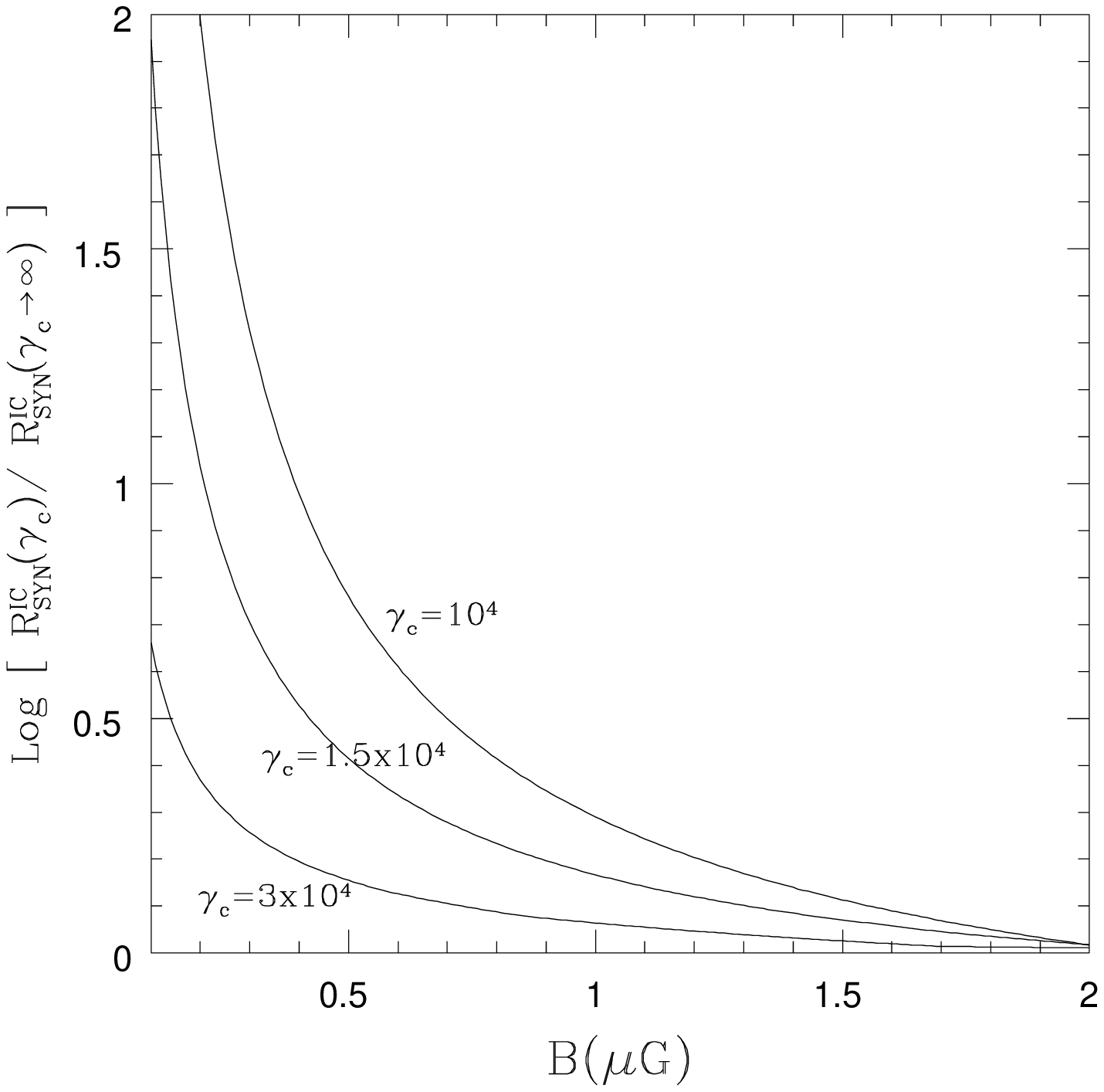}{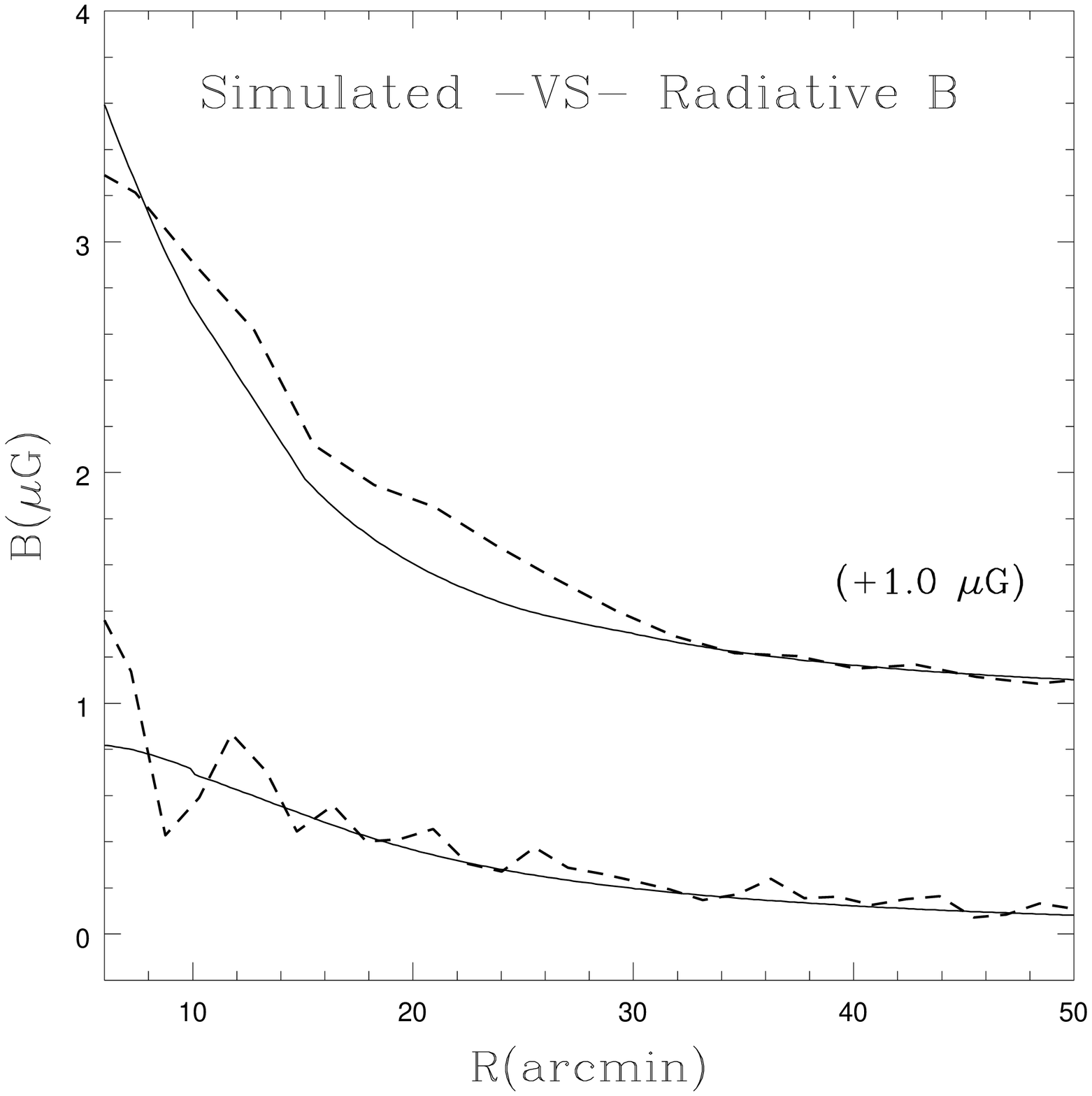}
\caption{
{\bf Panel a)}: The ratio between the HXR/SYN calculated
assuming a high energy cut--off in the electron spectrum
(given in the panel) and the HXR/SYN calculated with 
an infinite power law is given as a function
of $B$.
The SYN emission is calculated at 1.4 GHz.
{\bf Panel b)}: The comparison between the radial profiles
of $B$ in Coma derived
from the {\it two phase model} (solid lines) 
and from numerical simulations (dashed lines)
are reported in the case
of medium (upper) and low (lower) field.
For display purposes, the profiles in the case of medium
fields are shifted by 1.0$\mu G$.}
\end{figure}

We stress, however, that the result shown in Fig.5a might be 
misleading as the energy of the cut--off in the electron spectrum and
the value of the $B$ field are constrained by the shape 
of the radio spectrum.
A possibility to check how much the field discrepancy can be alleviated
is given by detailed calculations
based on models of radio halos whose parameters are forced to 
reproduce the overall radio synchrotron properties of the Coma
halo (i.e., brightness profiles, integrated radio spectrum,
and radial spectral steepening).
Given a model for particle acceleration, the comparison between 
the overall radio properties and the model expectations provides
a set of possible radial profiles of the cluster magnetic field.
In Fig.5b we report two representative profiles obtained
by fitting the radio properties with the {\it two--phase model}
(Brunetti et al. 2001a) compared with two profiles obtained
by independent numerical simulations (Dolag et al., 2002).
The IC calculations in the case of the low field model in Fig.5b
provide
a HXR flux from the Coma cluster compatible with the observations,
whereas IC scattering would account for $\sim 30$\% of the
observed HXR in the case of medium field profile.
It is important to notice that, while the volume averaged
magnetic field strength in both the cases in Fig.5b 
is $ \sim 0.3-0.4 \mu G$, the magnetic field strength in the 
cluster core region ($\leq 12$ arcmin), i.e. the region in
which RM observations are effective, is between
0.8 and 2 $\mu G$. This value is compatible within a factor
of $\sim 2$ with the values of the magnetic fields inferred
by RM observations in a number of clusters
(e.g., Clarke et al., 2001; Carilli \& Taylor, 2001).
In addition, when a power spectrum of the field is assumed,
the magnetic field strengths inferred
from RM data is lower than that inferred with standard
recipes
(e.g., Dolag et al., 2002; Newman, Newman, Rephaeli, 2002;
Govoni et al., this meeting).
If so, the discrepancy between IC and RM field may be considerably
alleviated.
The improvement in sensitivity provided by the
future X--ray observatories (e.g., ASTRO--E2, NEXT)
is impressive and it will probably allow us to
test the IC hypothesis.

\section{Conclusions}

Highly relativistic electrons (i.e., $\gamma > 10^3$)
can be injected in clusters of galaxies by several 
processes : they can be accelerated by merger shocks (Pop A),
they can be relic relativistic electrons 
(i.e., $\gamma \sim 10-100$) reaccelerated by cluster turbulence
(Pop B), and they can be secondary electrons injected by 
hadronic collisions (Pop C).
We examine three diagnostic which can help us to better 
understand the origin of the relativistic electrons producing
the observed radio synchrotron emission :

\begin{itemize}
\item[{\it i)}] Electrons accelerated by strong merger
shocks (Pop. A) cannot produce synchrotron emission
diffuse on $\geq$Mpc scale as that of classical radio halos.
This is due to the short radiative lifetime of the
electrons after being accelerated in the shock region.
A possibility to accomodate the Mpc sizes within the 
Pop. A is given by very fast (${\cal M} > 5$) - but unlikely -
shocks crossing the cluster center or by the presence
of cluster turbulence in addition to the merger shocks.  
\item[{\it ii)}] The comparison between the radio and
the soft X--ray brightness of a number of radio halos
indicates that the profile of the
radio emission is broader than that 
of the X--ray thermal emission. 
This appears to be difficult to be
accomodated within secondary models (Pop. C) which
would yield narrower radio profiles.
A possibility to skip this problem is to admit an
{\it ad hoc} increasing fraction of energy density 
of the relativistic protons with radius. 
However, at least in some cases,
this would imply an energetics of
the relativistic protons 
higher than that of the thermal pool.
\item[{\it iii)}] The spectral cut--off and 
radial spectral steepenings observed in the 
case of Coma (and in the mini--halo in the Perseus 
cluster) strongly point to the presence of
a cut--off in the spectrum of the emitting electrons.
This cut--off may be naturally accounted for 
if the synchrotron
emission is produced by reaccelerated (Pop.B) electrons,
whereas it is not expected in the case
of secondary electrons (Pop.C).
Future studies will clarify how much 
synchrotron spectral cut--offs and
radial steepenings are common in radio halos.
\end{itemize}

\noindent
Points i)--iii) would suggest that radio halos are powered 
by the synchrotron emission from
electrons reaccelerated (Pop.B) in the cluster volume during
merger events. This conclusion is, however, based on 
detailed studies of only few radio halos. 
As a consequence, detailed observations are still 
required to better 
understand the origin of radio halos.

The origin of the HXR emission detected in few clusters of
galaxies is still matter of debate.
It could be IC emission from relativistic electrons belonging to
the same population of electrons responsible for the large scale 
radio emission.
Alternatively HXR emission might result from bremsstrahlung
emission from a supra--thermal tail of electrons.
Both these hypothesis have problems: the IC emission would require
a magnetic field value in apparent
disagreement with that inferred from RM
observations,
while the supra--thermal bremsstrahlung requires a 
too large amount of energy if emitted for $> 10^8$yrs.
We have shown that the discrepancy between 
the field value obtained from the IC assumption and that from
the RM observations can be significantly reduced.
It is now clear that the combination of spatial trends and
inhomogenities in the thermal gas and magnetic field distribution 
with the presence of a cut--off in the electron spectrum at the 
energies of the radio emitting electrons would allow the IC magnetic 
field to be in better agreement with that from the RM observations.
Although these assumptions are {\it a posteriori} and might seem 
a sort of {\it conjuring tricks}, it should be noticed
that the presence of a high energy cut--off 
at $\gamma_{\rm c} \sim 10^{4}$ would really 
results from models of radio halos invoking 
the reacceleration of relic relativistic electrons. 
In addition a decreasing
radial profile of the magnetic field strength is naturally
expected.
In the framework of Pop.B models, we have shown that assuming 
the same model parameters
necessary to reproduce the general radio properties of the Coma
halo (brightness profile, integrated radio spectrum, and radial
spectral steepening), the resulting IC HXR would easily be about
$\sim 30-100$\% of that observed.
In this case, the resulting magnetic field strength in the cluster
core is of the order of $\sim 1 \mu G$ which is in rough agreement with
the RM observations especially when the field strength 
from the RM data is calculated assuming a power spectrum of the
field itself.

\acknowledgments
I would like to thank S.Bowyer, C.-Y. Hwang and the local
organizing committee for organizing such an enjoyable and
interesting conference. 
I am grateful to L.Feretti and P.Blasi
for useful discussions and to K.Dolag for providing
the magnetic field profiles reported in Fig.5b.

\end{document}